\documentclass[a4paper,11pt]{article}
\usepackage{jcappub} 
\usepackage{lineno}
\usepackage{amsmath,amssymb}
\usepackage{mathtools}
\usepackage{physics}
\usepackage{xcolor}
\usepackage{color, colortbl}
\usepackage{cancel}
\usepackage{multirow} 
\usepackage{booktabs} 
\usepackage{makecell} 
\usepackage{mathrsfs}
\usepackage{graphicx}
\usepackage[compat=1.1.0]{tikz-feynman}
\usepackage{enumerate}
\usepackage{inputenc}
\usepackage[shortlabels]{enumitem}
\usepackage{xspace}
\usepackage[normalem]{ulem}
\usepackage{slashed}

\definecolor{darkgreen}{RGB}{0,110,50}

\newcommand{\vmin}{v_{\mathrm{min}}}

\newcommand{\Nsig}{N_{\mathrm{sig}}}
\newcommand{\sigN}{\sigma_N}
\newcommand{\ER}{E_R}

\title{\boldmath A Freeze-In Interpretation of the LZ High-Energy Nuclear Recoil Event}

\author{D. Cabo-Almeida$^a$,}
\affiliation[a]{Institut f\"ur Theoretische Physik, Georg-August-Universit\"at G\"ottingen, Friedrich-Hund-Platz 1, 37077 G\"ottingen, Germany}

\author{F. Costa$^b$, }
\affiliation[b]{School of Theoretical Physics, Dublin Institute for Advanced Studies, 10 Burlington Road, Dublin, D04 C932, Ireland}
\author{D. Feiteira$^c$,}
\affiliation[c]{Department of Physics and Helsinki Institute of Physics, Gustaf Hällströmin katu 2a, FI-00014 Helsinki, Finland}

\author{V. Oliveira$^{d,e,f}$}
\affiliation[d]{Department of Physics, Lund University, SE-223 62 Lund, Sweden}
\affiliation[e]{Departamento de Física da Universidade de Aveiro, Campus de Santiago, 3810-183 Aveiro, Portugal}
\affiliation[f]{Laboratório de Instrumentação e Física Experimental de Partículas (LIP), Universidade do Minho, 4710-057 Braga, Portugal }

\emailAdd{david.caboalmeida@uni-goettingen.de}
\emailAdd{fcosta@stp.dias.ie}
\emailAdd{duarte.dasilvafeiteira@helsinki.fi}
\emailAdd{viniciuslbo@ua.pt}

\abstract{
The recent LUX-ZEPLIN (LZ) experiment analysis reported a high energy nuclear recoil candidate at $E_R \simeq 248~{\rm keV}$, motivating interpretations in terms of inelastic dark matter. We study this event in a pseudo-Dirac fermion model with a vector boson mediator $Z'$, assuming a low-reheating temperature and dark matter production through freeze-in at stronger coupling. In contrast to the thermal freeze-out case, the reheating temperature provides an additional parameter controlling the relic abundance and breaks the one-to-one relation between the dark matter mass and direct-detection cross section. As a result, the LZ candidate can be reproduced over a continuous region of parameter space, including mass splitting below the thermal benchmark. }

\begin{document}

\hfill DIAS-STP-26-20
\vspace{-1 cm}

\maketitle

\flushbottom

\section{Introduction}

The microscopic nature of dark matter (DM) remains one of the major open questions in modern physics. While its existence is firmly established by cosmological and astrophysical observations, its particle nature is still unknown. A leading paradigm is that of weakly interacting massive particles (WIMPs), whose relic abundance is set by thermal freeze-out from the Standard Model (SM) plasma. For interaction strengths around the electroweak scale, this mechanism can naturally reproduce the observed DM abundance, motivating an extensive direct detection program, including experiments such as LUX-ZEPLIN (LZ) and XENONnT~\cite{Arcadi:2024ukq}. 

A complementary production mechanism is freeze-in, in which DM never reaches thermal equilibrium with the SM bath because of sufficiently feeble interactions. Its relic abundance is instead built up gradually through rare production processes involving SM particles \cite{PhysRevLett.72.17,Hall:2009bx}. 
Freeze-in is commonly studied assuming a reheating temperature well above the DM mass $m_\chi$. In this regime, avoiding thermalization requires very small couplings between the dark and visible sectors, typically placing direct detection signals far beyond experimental reach. An alternative scenario, known as freeze-in at stronger coupling, is to consider a low-reheating temperature, $T_R \ll m_\chi$, for which DM production is Boltzmann suppressed. Reproducing the observed relic abundance then requires larger couplings, making the scenario potentially testable in direct detection experiments and at colliders~\cite{Khalil_2008,PhysRevD.109.075038,Cosme:2024ndc}.
More generally, varying $T_R$ allows freeze-in to extend continuously toward larger interaction strengths, filling the region in coupling space between the conventional feeble-coupling freeze-in regime and thermal freeze-out. In this sense, Boltzmann-suppressed freeze-in provides a framework to explore the full transition between non-thermal and thermal DM production. Low-reheating freeze-in has recently attracted growing attention as a way to extend freeze-in phenomenology toward larger interaction strengths and experimentally accessible signatures~\cite{bernal2026minimalfreezeindarkmatter,Bernal_2026,bernal2025freezingincannibalslowreheatingtemperature,koivunen2024probingsterileneutrinofreezein,Arcadi:2024obp,khan2026decayingvectordarkmatter,lee2024gravitymediateddarkmatterlow,Roy_2026,haghi2025probinghighreheatingtemperatures,Arias_2026,Henrich:2024rux,Feiteira:2026qme,Feiteira:2026ucz,Arcadi:2024wwg,Lebedev:2024mbj,Costa:2026qlo,barman2025leptoncolliderwindowreheating,Barman_2025,Borah_2025}.

The recent extended-energy analysis of the LZ experiment~\cite{akerib2026searchdarkmatterparticle} provides a particularly interesting target for this framework. LZ reports a single nuclear-recoil candidate, recorded on 16 June 2023, with
$E_R = 248 \pm 23\,{\rm (stat)} \pm 23\,{\rm (sys)}\,{\rm keV}$
in a $2.84$ tonne-year exposure, with a global significance of $2.6\sigma$. While this is insufficient to establish a DM signal, the unusually large recoil energy makes the event sufficiently distinctive to motivate dedicated model interpretations~\cite{DiMauro:2026ldr}.
A natural realization is provided by pseudo-Dirac DM coupled to a vector mediator $Z'$ \cite{Konar_2020}. In such models, a small Majorana mass splitting between the two nearly degenerate fermionic states leads to endothermic scattering \cite{garcia2025notsoinelasticdarkmatter,garcia2025openingparameterspacesubgev}, which suppresses low-energy recoils and can shift the spectrum toward the high-energy region relevant for the LZ event. The same $Z'$ framework has also been studied in the context of low-reheating freeze-in~\cite{Arcadi:2024obp}.
Thermal Higgsino DM has also been proposed as a possible explanation for the LZ candidate event \cite{DiMauro:2026ldr,yin2026pqsymmetrichighscalesusyinterpretation,bose2026goodnushiggsinodark,bisal2026higgsinodarkmatterinterpretation,du2026tevhiggsinointerpretationlz}. However, such a candidate would produce additional higher-energy recoils, which were not reported by the LZ collaboration \cite{Rodd:2026tyn}. Moreover, solar-capture constraints have been shown to rule out thermal Higgsinos as possible candidates \cite{dimauro2026solarcapturetestsinelastic,nguyen2026solarneutrinoconstraintsinelastic}. Other dark matter explanations have also been proposed~\cite{Heikinheimo:2026kwp,Zhu:2026dag,Lee:2026jxl,Lee:2026xxh,Okada:2026eol,Bandyopadhyay:2026gjw,Wang:2026ytg,Kotlarski:2026pep,Lee:2026wof,Unwin:2026rdp,Smirnov:2026aqk,das2026inelasticselfinteractingdarkmatter,borah2026inelasticsingletdoubletfermiondark,elahi2026vectorlikeleptoninterpretationhighenergy,jeesun2026atmosphericneutrinoupscatteringexplanation,sannino2026interpretinglz248kev,jung2026darkdiffractionlzscreened,ahmed2026kaluzakleindarkphotonmediationinelastic,ahmed2026xenonisotopefilteringkinematic,ahmed2026darkdimensionorigingeometricinelastic,langhoff2026sneakysneutrinoscatteringlz}.

In Ref.~\cite{DiMauro:2026ldr}, the LZ candidate is interpreted in a thermal pseudo-Dirac scenario in which the relic-density requirement fixes the interaction strength, and therefore the direct-detection cross section, for a given DM mass.
In Ref.~\cite{deLima:2026shq}, standard high temperature freeze-in was found to overproduce DM in order to explain the LZ candidate in a leptophobic $Z'$ model, while showing, for one reheating temperature benchmark, it could accommodate the LZ event for a Boltzmann suppressed freeze-in.
In this work, we focus on the low-reheating temperature regime, 
where that DM is produced through Boltzmann-suppressed freeze-in at stronger coupling. The reheating temperature $T_R$ then becomes an additional parameter controlling the relic abundance, breaking the one-to-one relation between the mass and the direct detection cross section. As a result, the LZ candidate can be accommodated along a continuous region of parameter space rather than at a single thermal-relic target.
Moreover, we take into account the effects of exothermic scattering and its consequences for reproducing the LZ candidate event. The exothermic scattering plays an important role in direct detection when a non-negligible amount of DM is given by the heaviest pseudo-Dirac candidate. This was originally discussed in Refs.~\cite{heeba2023inelasticfreezein,brahma2023resonantpseudodiracdarkmatter}. The hadrophilic exothermic analysis of the low-reheating freeze-in scenario was studied in Ref.~\cite{he2026hadrophilicinelasticfreezeindark}.

We distinguish between two pseudo-Dirac DM models: a leptophobic model for which the vector $Z'$ couples only to quarks, and a universal coupling model, allowing couplings between $Z'$ and all SM leptons. In the leptophobic model, the heaviest pseudo-Dirac candidate can be shown to be long lived for masses around the TeV scale and mass splitting in  the $100$ keVs range. Thus, exothermic scattering's effect on direct detection must be considered, as pointed out in Refs.~\cite{heeba2023inelasticfreezein,brahma2023resonantpseudodiracdarkmatter}. In the lepton universal coupling model, the depletion of the heaviest pseudo-Dirac is accomplish for mass splitting around $100$ keVs range and the process is fully dominated by endothermic scattering.

The paper is organized as follows. In Sec.~\ref{sec:model}, we introduce the pseudo-Dirac DM model and its leptophobic and universal-coupling realizations. In Sec.~\ref{sec:LZ_DD}, we discuss the direct-detection kinematics of the LZ candidate and define the one-event benchmark and compare the effects of universal and leptophobic couplings and of endothermic and exothermic scattering on direct detection. In Sec.~\ref{sec:LR}, we describe DM production through low-reheating-temperature freeze-in. In Sec.~\ref{Sec.:5}, we combine the relic-density and direct-detection requirements, present the numerical results for both coupling scenarios, and discuss their phenomenological implications. Finally, we conclude in Sec.~\ref{sec:concl}.

\section{Pseudo-Dirac Fermions}
\label{sec:model}

Pseudo-Dirac fermions provide a simple and well-motivated realization of inelastic dark matter~\cite{Tucker-Smith:2001myb,Hsieh2008,DeSimoneSanzSato2010,IzaguirreKrnjaicShuve2016,CarrilloGonzalez:2021lxm}. The basic idea is that a Dirac fermion, protected by an approximately conserved dark fermion number, is perturbed by a small Majorana mass term. The latter breaks the corresponding symmetry and splits the original Dirac state into two nearly degenerate Majorana mass eigenstates. The small mass splitting is therefore technically natural, since the exact Dirac limit restores the dark fermion number symmetry. The mass splitting provides a natural explanation for the inelastic scattering favoured by the LZ observed event.

We introduce two left-handed Weyl fermions, $\xi$ and $\eta$, which combine in the symmetry-preserving limit into the four-component Dirac field
\begin{align}
\Psi =
\begin{pmatrix}
\xi_\alpha \\
\eta^{\dagger\dot{\alpha}}
\end{pmatrix}.
\end{align}
The most general mass Lagrangian is
\begin{align}
\mathcal{L}_{\rm mass}
=
-m_D\,\xi\eta
-\frac{1}{2}m_L\,\xi\xi
-\frac{1}{2}m_R\,\eta\eta
+\mathrm{h.c.},
\end{align}
where $m_D$ denotes the Dirac mass while $m_L$ and $m_R$ are Majorana masses. In the pseudo-Dirac regime,
\begin{align}
m_L,m_R\ll m_D,
\end{align}
the spectrum consists of two approximately degenerate Majorana fermions, $\chi_1$ and $\chi_2$, with masses
\begin{align}
m_{\chi_1}
&\simeq
m_D-\frac{m_L+m_R}{2},
&
m_{\chi_2}
&\simeq
m_D+\frac{m_L+m_R}{2},
\end{align}
and hence
\begin{align}
\label{eq:gap}
\delta\equiv m_{\chi_2}-m_{\chi_1}
\simeq m_L+m_R \,.
\end{align}
We identify $\chi_1$ as the stable DM state and $\chi_2$ as its slightly heavier partner. Notice that the Dirac limit is continuously recovered for $m_L,m_R\rightarrow0$, in which case $\delta\rightarrow0$ and the dark fermion-number symmetry is restored.

We assume that the dark sector communicates with the Standard Model through a neutral vector mediator $Z'_\mu$. Before diagonalizing the fermion mass matrix, the relevant interactions can be written as
\begin{align}
\mathcal{L}_{\rm int}
\supset
g_\chi Z'_\mu\bar{\Psi}\gamma^\mu\Psi
+
g_f Z'_\mu\sum_f\bar f\gamma^\mu f \,,
\end{align}
where $f$ represents the SM fermions; $g_\chi$ and $g_f$ denote the mediator couplings to the dark and visible sectors, respectively. A characteristic feature of the pseudo-Dirac construction is that the vector interaction becomes predominantly off diagonal in the Majorana mass basis. In particular, in the symmetric limit $m_L=m_R$ one obtains
\begin{align}
\mathcal{L}_{\rm int}
\supset
i g_\chi Z'_\mu
\bar{\chi}_2\gamma^\mu\chi_1
+
g_f Z'_\mu\sum_f\bar f\gamma^\mu f \,.
\end{align}
The diagonal vector current of an individual Majorana fermion vanishes identically, $\bar{\chi}_i\gamma^\mu\chi_i=0$, so that the leading tree-level interaction converts one dark-sector state into the other. Direct detection therefore proceeds only through the endothermic transition
\begin{align}
\chi_1+N\longrightarrow\chi_2+N,
\label{eq:endo}
\end{align}
with an energy cost set by the mass splitting $\delta$, when $\chi_2$ is efficiently depleted after production, while exothermic scattering
\begin{align}
\chi_2+N\longrightarrow\chi_1+N,
\label{eq:exo}
\end{align}
is favoured over the endothermic when a non-negligible population of $\chi_2$ survives until today~\cite{heeba2023inelasticfreezein}.

For momentum transfers relevant to direct detection, the $Z'$ exchange can be described by a contact interaction provided that $m_{Z'}\gg 1\,\mathrm{GeV}$. Upon integrating out the mediator, we obtain
\begin{equation}
\mathcal{L}_{\mathrm{eff}}=i\frac{g_\chi g_q}{m_{Z'}^2}\left(\bar{\chi}_2\gamma^\mu\chi_1\right)\sum_q\bar q\gamma_\mu q .
\label{eq:DD_EFT}
\end{equation}

The sum in Eq.~\eqref{eq:DD_EFT} runs over all light SM quarks, since only they contribute to the nucleon matrix element at tree level. For universal quark couplings, the nucleon matrix elements of the vector current give
\begin{equation}
c_1^p=c_1^n\equiv c_1^s=\frac{3g_\chi g_q}{m_{Z'}^2}.
\label{eq:c1_matching}
\end{equation}
The interaction therefore reduces to the non-relativistic operator
$\mathcal O_1=\mathbf 1_\chi\mathbf 1_N$ with an isoscalar coupling~\cite{Fan:2010gt,Fitzpatrick:2012ix,Anand:2013yka},
corresponding to the inelastic $\mathcal O_1^s$ benchmark considered
by LZ. The associated DM--nucleon cross section is
\begin{equation}
\sigma_N=\frac{\mu_N^2}{\pi}\left(\frac{3g_\chi g_q}{m_{Z'}^2}\right)^2,
\label{eq:cross_sectionN}
\end{equation}
where $\mu_N=\frac{m_{\chi_1}m_N}{m_{\chi_1}+m_N}$ is the reduced mass of the nucleon. Thus, direct detection probes the combination $g_\chi g_q/m_{Z'}^2$, which will later be confronted with the DM production requirement.

The coupling to leptons does not enter the DM-nucleon cross section at tree level, since only the quarks contribute to the nucleon matrix element. It does, however, control the $Z'$ width, the annihilation channels available in the early Universe, the collider bounds on the mediator and the $\chi_2$ depletion. 
We therefore consider two versions of the model. In the \emph{leptophobic} case the $Z'$ couples only to the dark sector and to the quarks, so that $g_f = g_q$ and $g_\ell = 0$. In the universal coupling
case it couples universally to all fermions, so $g_f=g_q=g_l$. 
We keep both throughout and compare them in Sec.~\ref{Sec.:5}.

\section{Direct Detection Physics of the LZ event}
\label{sec:LZ_DD}
LZ reports one nuclear-recoil candidate with reconstructed energy
\begin{equation}
E_R = 248 \pm 23\,({\rm stat}) \pm 23\,({\rm sys})~{\rm keV}
\end{equation}
in an exposure of $2.84~{\rm tonne\,yr}$ \cite{akerib2026searchdarkmatterparticle}. Although an elastic TeV-scale WIMP can kinematically produce such a recoil, the standard elastic spin-independent (SI) spectrum decreases rapidly with recoil energy. In particular, the momentum transfer associated with
the candidate is
\begin{equation}
q=\sqrt{2m_AE_R}\simeq 0.25~{\rm GeV}
\end{equation}
for xenon, where the coherent nuclear response is already strongly suppressed by the form factor. Consequently, normalizing an elastic-SI spectrum to produce one event near $248~{\rm keV}$ also predicts a much larger population at lower recoil energies. Quantitatively, one event in the interval $215<E_R/{\rm keV}<300$ would correspond to $\mathcal O(10^3)$ events between $5.4$ and $55~{\rm keV}$~\cite{DiMauro:2026ldr}. This strong mismatch between the predicted low- and high-energy populations makes conventional elastic-SI scattering an unlikely explanation of the isolated high-energy candidate, although a definitive assessment requires the full LZ likelihood. Let us discuss two distinct cases: when the direct detection signal is dominated by the endothermic scattering, Eq.~\eqref{eq:endo},
and the exothermic scattering, Eq.~\eqref{eq:exo}.

\paragraph{Endothermic scattering}
The mass gap introduced in Eq.~\eqref{eq:gap} modifies the minimum incoming DM velocity required to produce a nuclear recoil of energy $E_R$~\cite{Tucker-Smith:2001myb,Barello:2014uda}
\begin{equation}
v_{\rm min}^{\rm endo}(E_R)=
\frac{1}{\sqrt{2m_AE_R}}\left(    \frac{m_AE_R}{\mu_{\chi A}}+\delta\right).
\label{eq:vmin_inelastic}
\end{equation}
For $\delta>0$, part of the incoming kinetic energy is used to produce the heavier state. At low recoil energies, the velocity threshold is dominated by this excitation energy, whereas at high recoil energies it is dominated by the energy transferred to the nucleus. The competition between these two effects gives a minimum at

\begin{equation}
\label{eq:ERstar}
E_R^\star=\frac{\mu_{\chi A}}{m_A}\,\delta .
\end{equation}
Therefore, $E_R^\star$ is the recoil energy corresponding to the smallest possible incoming velocity and is accessible to the largest fraction of the DM halo. It provides the characteristic recoil scale of inelastic scattering. Taking the LZ candidate to lie close to this scale suggests
\begin{equation}
\label{eq:delta_estimate}
\delta \simeq \frac{m_A}{\mu_{\chi A}}\,(248~{\rm keV}).
\end{equation}
\paragraph{Exothermic scattering} If the excited state $\chi_2$ survives until today with a non-negligible abundance, the reverse transition
$\chi_2+N\to\chi_1+N$
is also allowed. In this case the mass splitting is released into the final state, and the minimum incoming velocity becomes
\begin{equation}
    v_{\min}^{\rm exo}(E_R)=
    \frac{1}{\sqrt{2m_AE_R}}
    \left|\frac{m_AE_R}{\mu_{\chi A}}-\delta\right| .
\end{equation}
It vanishes at the same characteristic recoil energy, Eq.~\eqref{eq:ERstar},
where the energy released in the transition matches the nuclear recoil energy. Therefore, recoils near $E_R^\star$ can be produced by the full halo population, rather than only by its high-velocity tail. The exothermic spectrum is therefore kinematically focused around this energy, while recoils far below or above $E_R^\star$ require increasingly large incoming velocities. In particular, exothermic scattering still suppresses sufficiently low-energy recoils, since $v_{\min}^{\rm exo}\propto E_R^{-1/2}$ as $E_R\to0$. When the present-day $\chi_2$ abundance is comparable to that of $\chi_1$, this channel generally dominates over the endothermic one because it does not require an upscattering threshold. \\

The differential scattering rate per unit detector mass is
\begin{equation}
\frac{dR}{d\ER}=\frac{\rho_\chi}{m_\chi}\frac{\sigN}{2\mu_N^2}\sum_i\xi_i A_i^2F_i^2(\ER) \left[(1-f)\eta\left(\vmin^{(i) }{}^{\rm endo}\right)+f \eta\left(v_{\rm min}^{(i) }{}^{\rm exo}\right)\right],
\label{eq:rate}
\end{equation}
where the sum runs over the naturally occurring xenon isotopes, $\xi_i$ denotes their mass fractions, $\mu_N$ is the DM--nucleon reduced mass, $\sigma_N$ is the isoscalar DM--nucleon scattering cross section, and $f$ is the fraction of $\chi_2$ of the total DM in the halo. Assuming equal initial abundances of $\chi_1$ and $\chi_2$, we estimate this fraction as $f=\frac{1}{2}e^{-t_{\rm uni}/\tau_{\chi_2}}$, where $\tau_{\chi_2}$ is the lifetime of the excited state and $t_{\rm uni}$ is the age of the Universe. We model the coherent nuclear response using the Helm form factor~\cite{Lewin:1995rx}, evaluated separately for each isotope. The mean inverse speed is 
\begin{equation}
\eta(\vmin)=
\int_{v>\vmin}d^3v\,
\frac{f_{\rm lab}(\mathbf v)}{v}.
\end{equation}
We adopt the Standard Halo Model, assuming a Maxwell--Boltzmann velocity distribution in the Galactic frame, shifted to the laboratory frame, with parameters $\rho_\chi=0.3~{\rm GeV\,cm^{-3}}$, $v_0=238~{\rm km\,s^{-1}}$ and $v_{\rm esc}=544~{\rm km\,s^{-1}}$~\cite{Baxter:2021pqo}. The Earth's speed in the Galactic frame is approximated to
\begin{equation}
v_E(t)=\overline v_E+\Delta v_E\cos\left[\frac{2\pi(t-t_0)}{T}\right],
\end{equation}
where $\overline v_E=250.2~{\rm km\,s^{-1}}$ is the time-averaged Earth speed, $\Delta v_E=14.9~{\rm km\,s^{-1}}$is the annual modulation amplitude, $T=1~{\rm yr}$, and $t_0$ corresponding to June 2. The expected number of detected events is obtained by folding the recoil spectrum with the LZ nuclear-recoil efficiency~\cite{akerib2026searchdarkmatterparticle}
\begin{equation}
\label{eq:nsig}
\Nsig=\mathcal{E}\int_{\ER^{\rm min}}^{\ER^{\rm max}}d\ER\,\varepsilon(\ER)\frac{dR}{d\ER} \,,
\end{equation}
where $\mathcal{E} = 2.84~{\rm tonne\,\cdot\,yr}$ is the exposure of the LZ, $\varepsilon(\ER)$ is the nuclear-recoil signal efficiency as a function of the true recoil energy, $dR/d\ER$ is the differential rate of Eq.~\eqref{eq:rate} and the minimum and maximum recoil energy are selected to be the limits of the region of interest by the experiment.

Given the presence of only one event, we do not perform a statistical fit to the LZ data. Instead, we use an order-one predicted event count as a necessary condition for the model to provide a plausible interpretation of the event. We therefore impose the central normalization
\begin{equation}
N_{\rm th} \simeq N_{\rm obs}-N_{\rm bkg} = 1-0.0106 \simeq0.989,
\label{eq:one_event_condition}
\end{equation}
where $N_{\rm bkg}=0.0106$ is the background expectation quoted for the high-$S1$ region containing the event. Since this number does not represent the complete event-level LZ background likelihood, Eq.~\eqref{eq:one_event_condition} should be regarded only as a one-event benchmark. It identifies parameter points capable of producing the observed candidate, but it does not define a best fit or a confidence region.


\section{Low-Reheating Temperature Freeze-in}
\label{sec:LR}
In Ref.~\cite{DiMauro:2026ldr}, the constraints derived in the previous section from the LZ candidate event were combined with the requirement of a thermal relic abundance. In this Section we move beyond thermalisation-only DM candidate and also consider the non-thermal production of $\chi$. We consider a low-reheating temperature, $T_R \ll m_{\chi_1}$, and produce the DM by freeze-in from the SM bath. Production from the bath therefore proceeds exclusively through $f \bar f \to \chi_1 \chi_2$. However, the $Z^\prime$ remains in thermal equilibrium with the bath through its coupling to the light fermions~\cite{Arcadi:2024obp}, so channels with an on-shell mediator are available in principle: the decay $Z' \to \chi_1\chi_2$, and the $t$-channel process $Z'Z' \to \chi_1\chi_1$. We have computed their relative contributions with \texttt{micrOMEGAs}~\cite{Belanger:2026asz} and find that the decay matters only around $m_{Z'} \simeq 2m_{\chi_1}$, where it briefly dominates, while $Z'Z'$ contributes less than $10\%$ of the total production. For the benchmarks considered below, $f\bar f \to \chi_1\chi_2$ accounts for essentially all of the production.

Let us here derive an analytical estimate for the DM production and the direct detection cross-section. Notice that the numerical results we will present in the next Section do not rely on any of the following assumptions and solve the general Boltzmann equation including the $Z'$ propagator and all tree-level contributions with the use of \texttt{micrOMEGAs}. Since $\chi_2$ may eventually decay into $\chi_1$, the relevant quantity is the total density $n = n_{\chi_1} + n_{\chi_2}$. Adding the Boltzmann equations of the two states cancels the decay terms. Moreover, in the freeze-in regime the DM density stays far below its equilibrium value, so the coannihilation rate term, $\Gamma_{\chi_1 \chi_2 \to q \bar q }$, is negligible. One is left with
\begin{equation}
\dot{n} + 3H n = 2\, \Gamma_{f \bar f \to \chi_1 \chi_2  }\,,
\label{eq:boltzmann}
\end{equation}
where the factor of $2$ accounts for the two DM particles, one $\chi_1$ and one $\chi_2$, produced in each $f\bar f \to \chi_1 \chi_2$ event. Therefore, in the absence of efficient processes that deplete or convert the excited state, their comoving abundances remain approximately equal, and $\chi_2$ survives as an appreciable component of the DM density until the present day. In the radiation-dominated era, the Hubble rate reads $H = \sqrt{g_{\rho} \pi^2/90}\, T^2/M_{\rm Pl}$, where $g_{\rho}$ counts the SM degrees of freedom (d.o.f.) in the energy density and $M_{\rm Pl} = 2.43 \times 10^{18}$ GeV is the reduced Planck mass. The production rate, $\Gamma_{f \bar f \to \chi_1 \chi_2  }$, can be written as the thermal coannihilation rate of the DM states into SM quanta~\cite{Arcadi:2024obp},
\begin{equation}
\Gamma_{f \bar f \to \chi_1 \chi_2  } = \Gamma^{\rm therm}_{\chi_1 \chi_2 \to f \bar f} = 2^2 \times \frac{T}{32 \pi^4} \int_{4m_{\chi_1}^2}^\infty ds \, \sigma \left(s - 4 m_{\chi_1}^2\right) \sqrt{s} \, K_1\left(\frac{\sqrt{s}}{T}\right),
\label{eq:interaction_rate}
\end{equation}
where the factor $2^2$ restores the internal degrees of freedom of the two initial states.

We are interested in  the heavy mediator limit, $m_{Z'}^2 \gg s$, hence the $Z^\prime$ mass can be integrated out for practical purposes and taking $m_f \to 0$, the cross section reads
\begin{equation}
\sigma \left( \chi_1 \chi_2 \to f \bar f \right) = \frac{N_c}{12 \pi}  \frac{g_f^2 g_\chi^2\sqrt{s} \left(s + 2 m_{\chi_1}^2 \right)}{m_{Z'}^4 \sqrt{s - 4m_{\chi_1}^2}}\,,
\label{eq:cross_section}
\end{equation}
whose $s$-wave contribution, obtained at $s \to 4m_{\chi_1}^2$, is
\begin{equation}
\langle \sigma v \rangle_{s\text{-wave}} = \frac{N_f\, g_f^2 g_\chi^2 m_{\chi_1}^2}{\pi\, m_{Z'}^4}\,,
\end{equation}
in agreement with Eq.~(56) of Ref.~\cite{DiMauro:2026ldr}. Here $N_f$ collects the multiplicity of the final states accessible to the mediator. In the universal coupling model the charged leptons contribute three further channels and the neutrinos, being purely left-handed, count as one half each, giving $N_f = 22.5$. Below each heavy-fermion threshold the corresponding channel closes and $N_f$ decreases accordingly.

For $T_R \ll m_{\chi_1}$ the production is Boltzmann suppressed. Expanding the Bessel function in Eq.~\eqref{eq:interaction_rate} for large argument, one finds
\begin{equation}
\Gamma^{\rm therm}_{\chi_1 \chi_2 \to f \bar f} \simeq \frac{N_f\, g_f^2 g_\chi^2 m_{\chi_1}^5 T^3}{2 \pi^4 m_{Z'}^4} e^{-\frac{2m_{\chi_1}}{T}}\,.
\end{equation}
Setting $N_f = 1$, this reproduces Eq.~(2.9) of Ref.~\cite{Arcadi:2024obp} under the substitution $g_f g_\chi \to V_f V_\chi$. The production is dominated by the initial moments after reheating. Integrating Eq.~\eqref{eq:boltzmann} from $T = T_R$ in terms of the yield $Y \equiv n/s_{\rm SM}$, where $s_{\rm SM} = (2\pi^2/45) g_s T^3$ is the SM entropy density and $g_s$ counts the entropic d.o.f., gives
\begin{equation}
Y \simeq \frac{135 \sqrt{10}}{4} \frac{N_f\, g_\star^{1/2}}{g_s^2} \frac{g_f^2 g_\chi^2 M_{\rm Pl}\, m_{\chi_1}^4}{\pi^7 m_{Z'}^4 T_R} e^{-\frac{2m_{\chi_1}}{T_R}}\,,
\end{equation}
where $g_*^{1/2} = \frac{g_s}{g_\rho^{1/2}}\left( 1 + T \frac{dg_s/dT}{3 g_s} \right)$. Using Eq.~\eqref{eq:cross_sectionN}, this expression can be rewritten in terms of the SI nucleon cross section as
\begin{equation}
Y \simeq \frac{15 \sqrt{10}}{4} \frac{N_f\, g_\star^{1/2}}{g_s^2} \frac{M_{\rm Pl}\, m_{\chi_1}^4\, \sigma_N}{\pi^6 T_R \mu_N^2} e^{-\frac{2m_{\chi_1}}{T_R}}\,.
\end{equation}
Imposing the observed relic abundance, $Y = 4.4 \times 10^{-10}\, {\rm GeV}/m_{\chi_1}$, one obtains a condition on the SI nucleon cross section,
\begin{equation}
\sigma_N = \frac{6.30 \times 10^{-51}\, {\rm cm}^2}{N_f} \frac{\mu_N^2}{m_{\chi_1}^2} \frac{T_R}{m_{\chi_1}} \frac{{\rm GeV}^2}{m_{\chi_1}^2} e^{\frac{2m_{\chi_1}}{T_R}}\,,
\label{eq:cross_section_final}
\end{equation}
where we take $g_\rho = g_s = 106.75$. Eq.~\eqref{eq:cross_section_final} is accurate over most of the parameter space, and departs from the numerical result only where the mediator sits close to the production threshold or where the coupling is large enough for the inverse process to matter.

For a given DM mass, Eq.~\eqref{eq:cross_section_final} gives the SI nucleon cross  section required to reproduce the observed abundance as a function of the reheating temperature. In Ref.~\cite{DiMauro:2026ldr}, an analogous condition is obtained for freeze-out DM. There, a fixed mass imposes a fixed SI nucleon cross section. In low temperature freeze-in this is no longer the case. The production rate is exponentially suppressed, so a lower reheating temperature requires a stronger coupling to compensate the suppression and still produce enough DM. Since $T_R$ is a free parameter, multiple SI nucleon cross sections reproduce the correct abundance for a fixed mass, and the LZ candidate event is reproduced along a continuous line in the $(\sigma_N, \delta)$  parameter space.

\section{Matching the LZ event with Freeze-in}
\label{Sec.:5}

In this Section, we investigate how the pseudo-Dirac fermion model introduced above can account for the LZ high-energy nuclear-recoil candidate when produced via freeze-in. 
First, we will discuss the leptophobic, and secondly the universal coupling scenario. Lastly, we will present their phenomenological implications.
The results are summarised in Table~\ref{tab:LZ_RD_ranges}.

\begin{table*}[t]
\centering
\caption{Ranges of the reheating temperature $T_R$, mass splitting $\delta$, characteristic recoil energy $E_R^\star$, and minimum velocity $v_{\rm min}^\star$ for points that simultaneously reproduce the observed DM relic abundance through low-temperature freeze-in and yield one expected LZ event. Only the portions lying within the LZ $90\%$ C.L. region are shown.}
\label{tab:LZ_RD_ranges}
\begin{tabular}{lcccccc}
\toprule
Scenario
& $m_{\chi_1}$ [TeV]
& $m_{Z'}$ [TeV]
& $T_R$ [GeV]
& $\delta$ [keV]
& $E_R^\star$ [keV]
& $v_{\rm min}^{\star,\rm endo}$ [$\mathrm{km\,s^{-1}}$] \\
\midrule
\multirow{5}{*}{Leptophobic}
& $0.5$ & $4.0$ & $[23,26]$ & $[26,350]$ & $[21,281]$ & $[219,801]$ \\
& $1.0$ & $6.25$ & $[42,47]$ & $[25,350]$ & $[22,312]$ & $[204,761]$ \\
& $2.5$ & $10.0$ & $[96,106]$ & $[25,350]$ & $[24,334]$ & $[197,735]$ \\
& $5.0$ & $15$ & $[190,195]$ & $[25,280]$ & $[25,273]$ & $[195,650]$ \\
& $10.0$ & $25$ & $[359,375]$ & $[25,115]$ & $[25,113]$ & $[194,414]$ \\
\midrule
\multirow{5}{*}{\shortstack{Universal\\coupling}}
& $0.5$ & $4.0$ & $[20,26]$ & $[26,303]$ & $[21,243]$ & $[219,745]$ \\
& $1.0$ & $6.25$ & $[39,47]$ & $[25,290]$ & $[22,258]$ & $[204,692]$ \\
& $2.5$ & $10.0$ & $[98,105]$ & $[25,214]$ & $[24,204]$ & $[197,575]$ \\
& $5.0$ & $15$ & $[194,195]$ & $[25,148]$ & $[25,145]$ & $[195,473]$ \\
& $10.0$ & $25$ & $[359,371]$ & $[25,83]$ & $[25,82]$ & $[194,352]$ \\
\bottomrule
\end{tabular}
\end{table*}

\subsection{Leptophobic Model}
\begin{figure}[t!]
    \centering
    \includegraphics[width=0.49\textwidth]{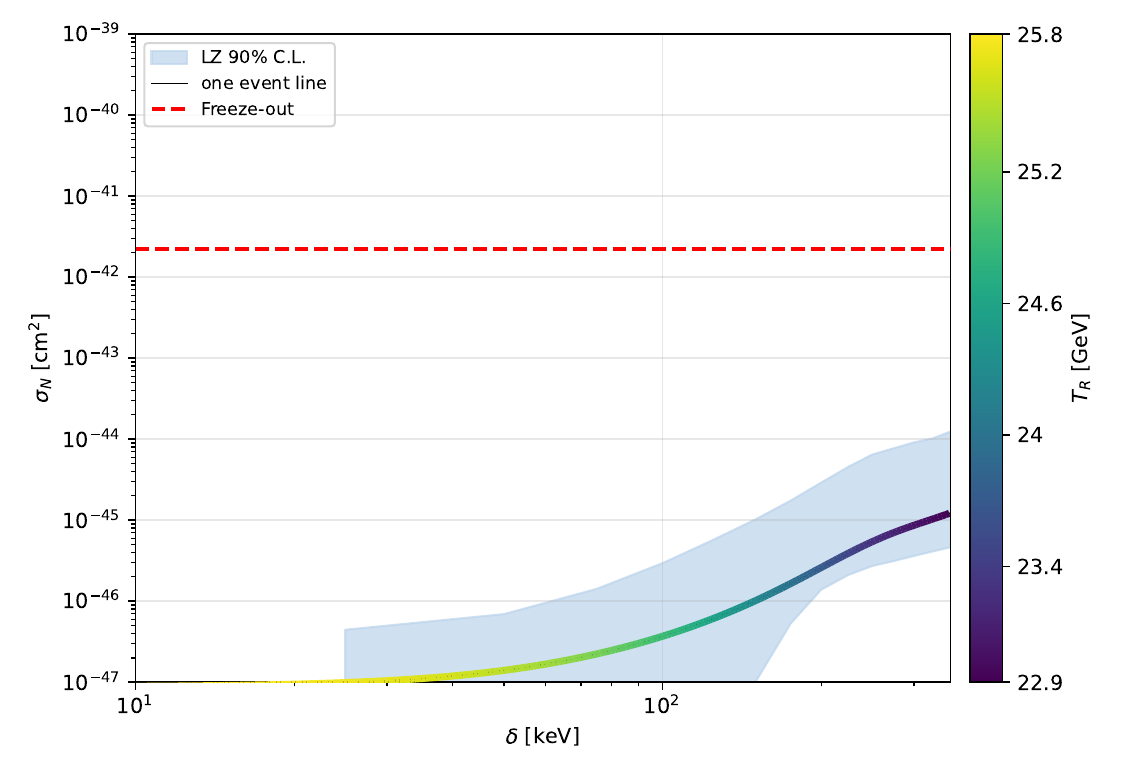}
    \includegraphics[width=0.49\textwidth]{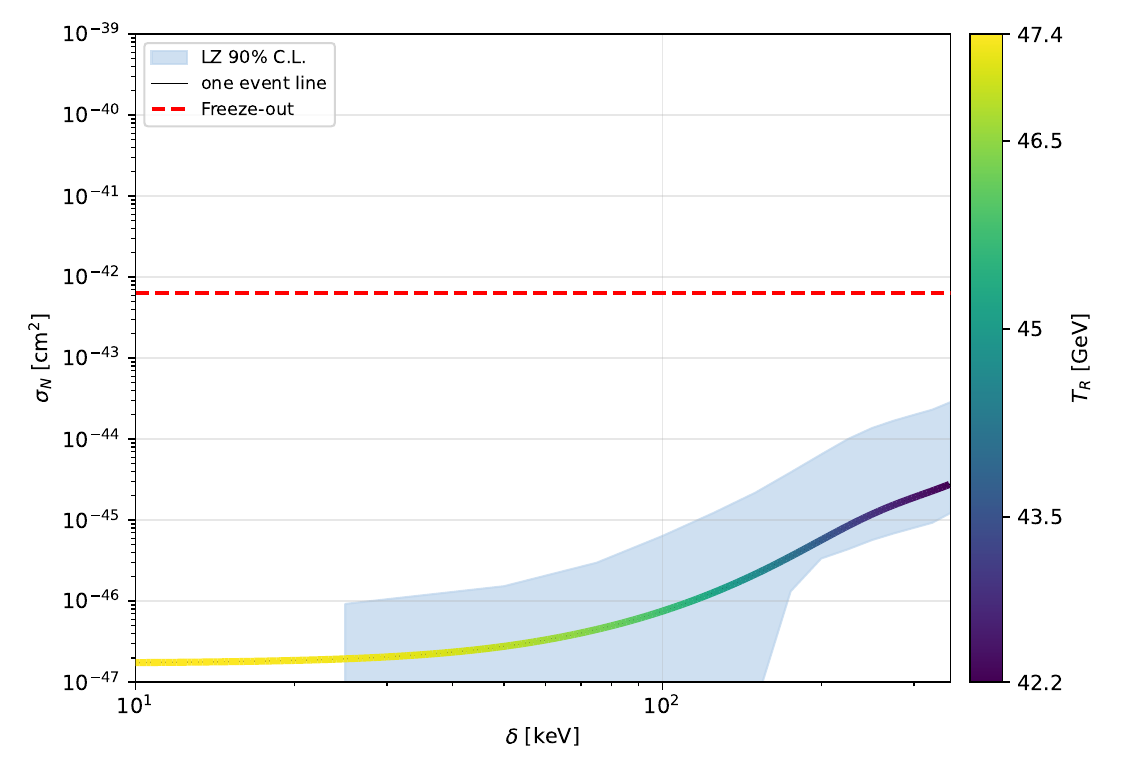}
    \includegraphics[width=0.49\textwidth]{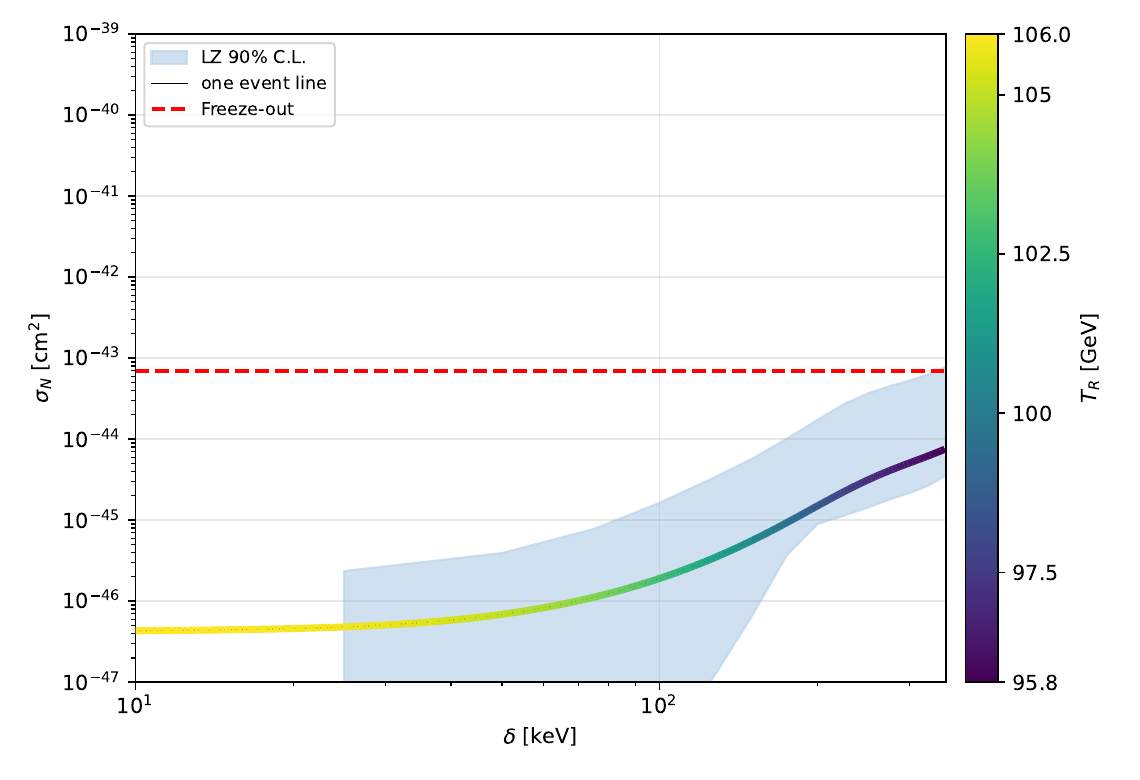}
    \includegraphics[width=0.49\textwidth]{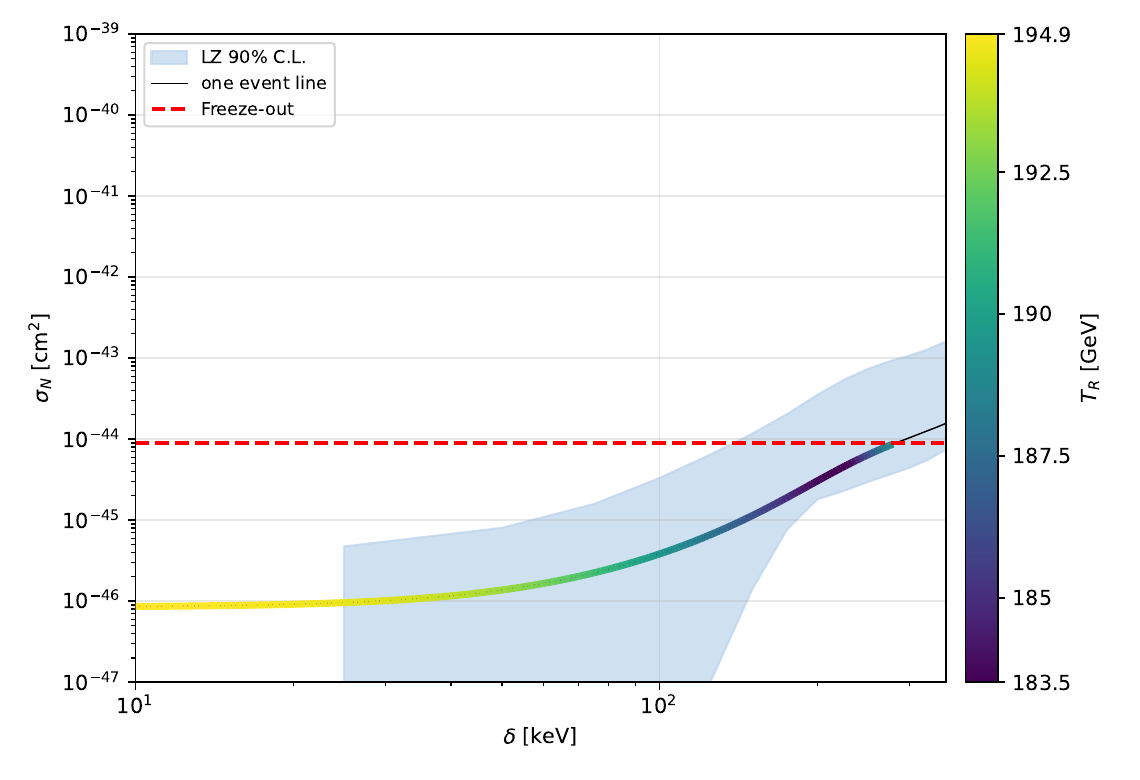}
    \caption{Comparison between the spin-independent DM--nucleon cross section \(\sigma_N\) that satisfies $\Omega h^2 =0.12$ and the one-event line that reproduces the LZ high-energy nuclear-recoil candidate, shown as a function of the pseudo-Dirac mass splitting \(\delta\), in the leptophobic scenario. The panels correspond to $(m_{\chi_1},m_{Z'})=(0.5,4)~{\rm TeV}$ (top left), $(1,6.25)~{\rm TeV}$ (top right), $(2.5,10)~{\rm TeV}$ (bottom left) $(5,15)~{\rm TeV}$ (bottom right). The blue band denotes the LZ $90\%$ C.L. region obtained for $f = 1/2$ as in Ref.~\cite{oliveira2026}, while the black curve gives the cross section required to produce one event. The red dashed line shows the thermal freeze-out prediction. The colored points reproduce both the observed DM abundance through low-temperature freeze-in and the one-event condition, with the color indicating the corresponding reheating temperature $T_R$.}
    \label{fig:leptophobic}
\end{figure}

The numerical results for the leptophobic scenario are shown in Fig.~\ref{fig:leptophobic}. In this scenario $\chi_2$ survives, so the LZ $90\%$ C.L. region is obtained for a halo with equal fractions of $\chi_1$ and $\chi_2$, following the recast of Ref.~\cite{oliveira2026}.
For each benchmark DM and mediator mass, we solve the full Boltzmann equation with
\texttt{micrOMEGAs} and determine the couplings $g_q$, and hence $\sigma_N$, that reproduces
the observed relic abundance as the reheating temperature is varied, while keeping $g_{\chi} =1$. As anticipated
from Eq.~\eqref{eq:cross_section_final}, decreasing $T_R$ increases the interaction strength required to
compensate for the Boltzmann suppression of DM production. The relic-density
condition therefore spans a continuous range of $\sigma_N$, rather than selecting the
single cross section characteristic of thermal freeze-out.

Fig.~\ref{fig:leptophobic} illustrates this behaviour for $(m_{\chi_1},m_{Z'})=(0.5,4)~{\rm TeV}$ (top left), \\ 
$(1,6.25)~{\rm TeV}$ (top right),  $(2.5,10)~{\rm TeV}$ (bottom left) and $(5,15)~{\rm TeV}$ (bottom right).
Along the one-event line, different values of the mass splitting can reproduce both
the observed relic abundance and the LZ candidate by adjusting $T_R$. In particular,
for $m_{\chi_1}=1~{\rm TeV}$ the allowed reheating temperatures are approximately
$T_R\simeq 42$--$47~{\rm GeV}$, while for $m_{\chi_1}=2.5~{\rm TeV}$ they increase to
$T_R\simeq 96$--$106~{\rm GeV}$. The available ranges are summarised in Table~\ref{tab:LZ_RD_ranges}. Notice that accounting for exothermic scattering renders the $m_{\chi} = 1 $ TeV thermal freeze-out interpretation of the LZ event incompatible with the data unless an additional mechanism in the UV completion efficiently depletes the excited state~\cite{DiMauro:2026ldr}. Low-reheating freeze-in provides the only solution for the leptophobic scenario with masses approximately below $5$ TeV. We notice that there is a continuous range of $\sigma_N$ that reproduces the relic density and sits on the one-line event. In fact, lower $\delta$ require lower $\sigma_N$ to satisfy the one-event line and that can be achieved by a higher reheating temperature. Among benchmark studied only $m_{\chi_1}=5 $ TeV can successfully reproduce the one-event signal and the relic abundance via thermal freeze out. Let us notice that lower values of $\delta$,
which would not reproduce the event in the thermal relic scenario, become now
viable with low-reheating freeze-in. Furthermore, we notice that the thermal cross section is an upper bound on the freeze-in cross section as standard in low-reheating scenarios.

\subsection{Universal Coupling Model}

\begin{figure}[t]
    \centering
    \includegraphics[width=0.49\textwidth]{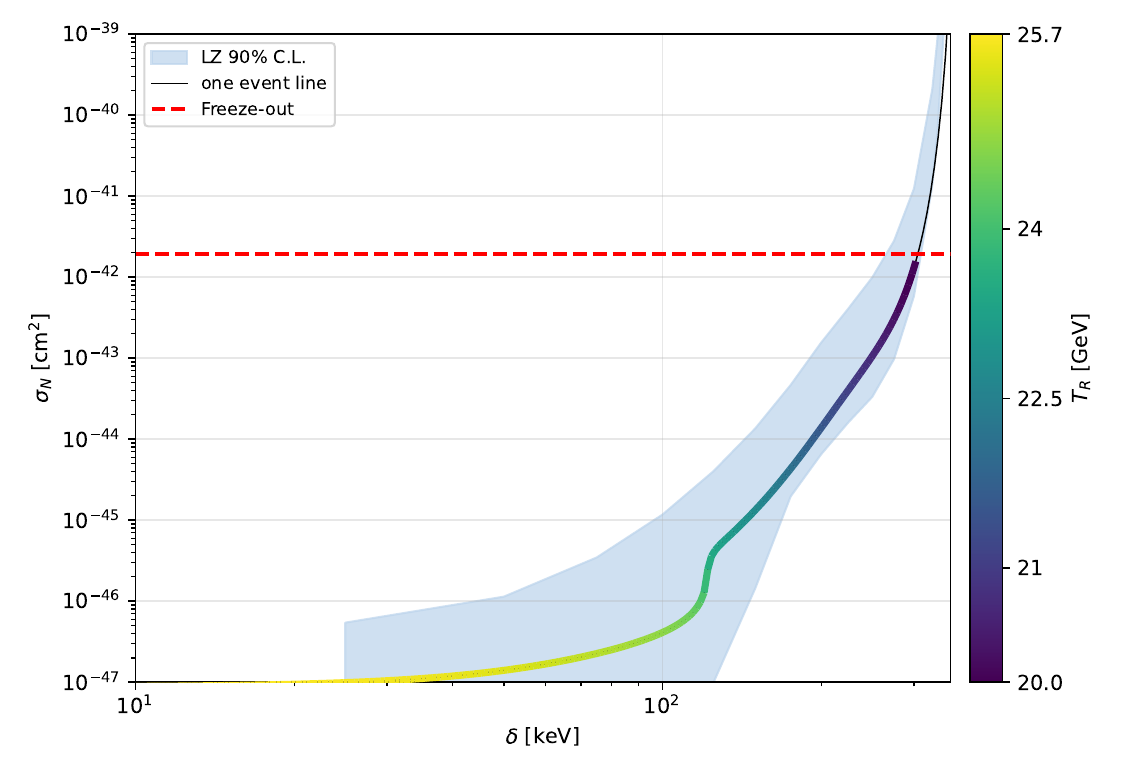}
    \includegraphics[width=0.49\textwidth]{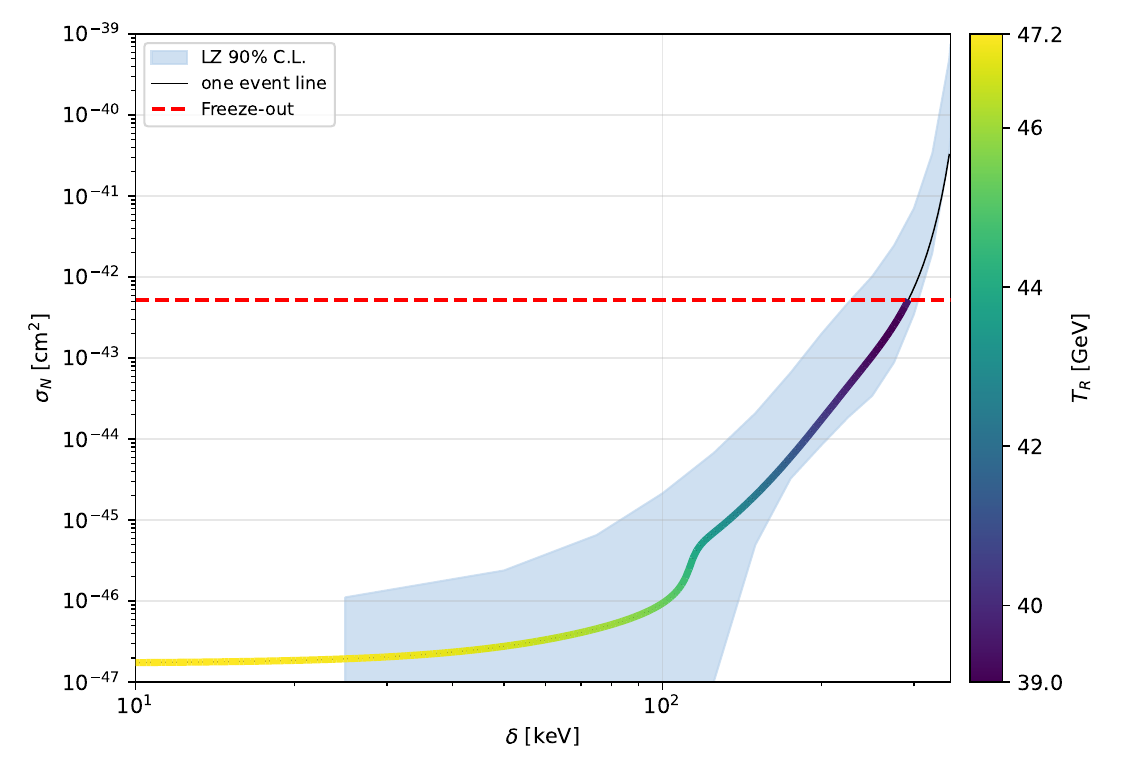}
    \includegraphics[width=0.49\textwidth]{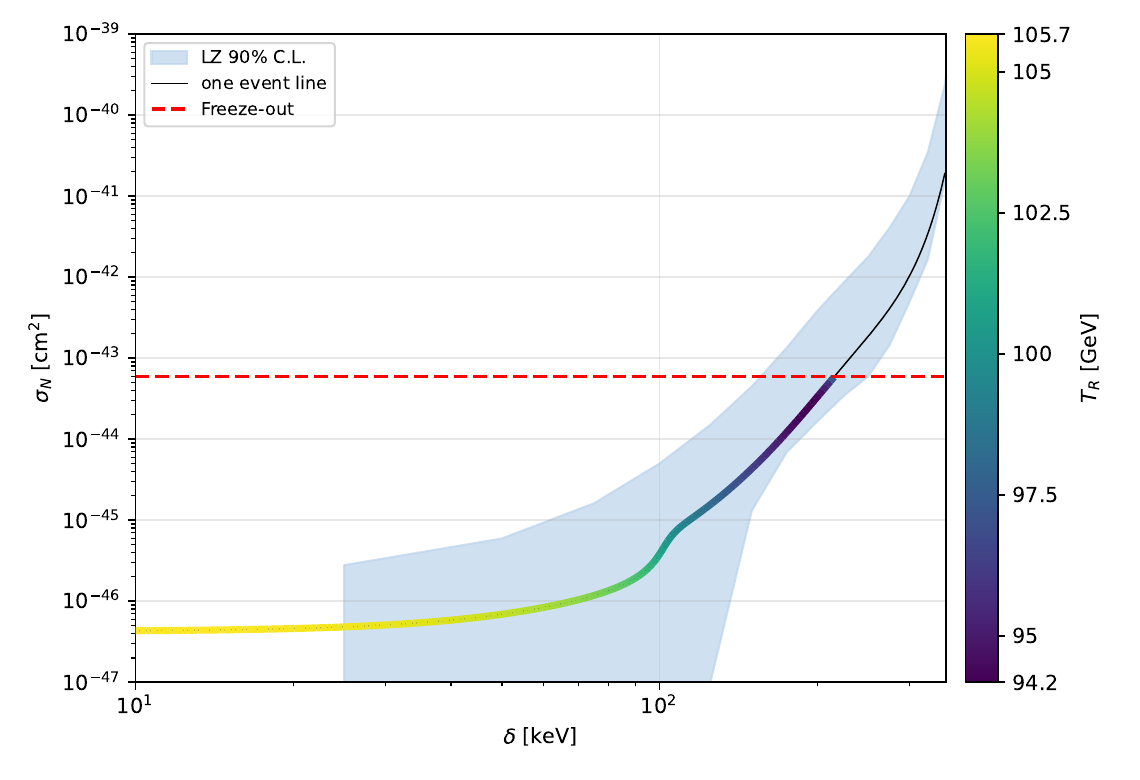}
    \includegraphics[width=0.49\textwidth]{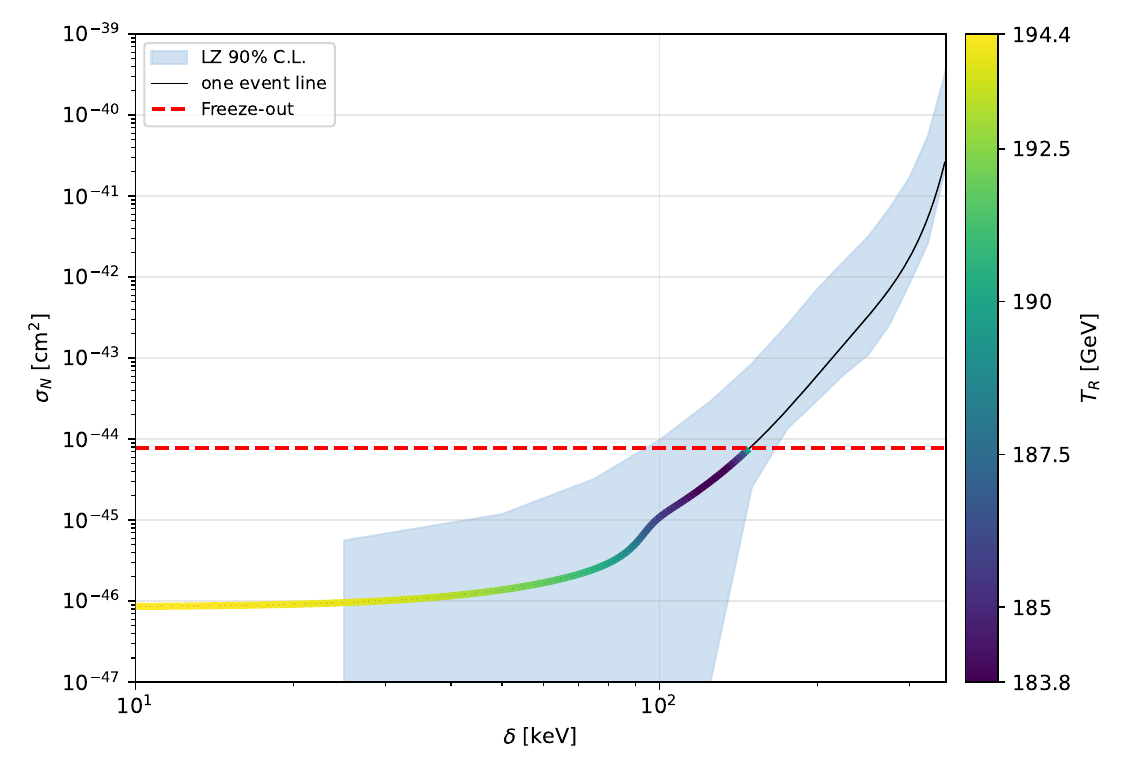}
    \caption{Comparison between the spin-independent DM--nucleon cross section \(\sigma_N\) that satisfies $\Omega h^2 =0.12$ and the one-event line that reproduces the LZ high-energy nuclear-recoil candidate, shown as a function of the pseudo-Dirac mass splitting \(\delta\), in the universal coupling scenario. The panels correspond to $(m_{\chi_1},m_{Z'})=(0.5,4)~{\rm TeV}$ (top left), $(1,6.25)~{\rm TeV}$ (top right), and $(2.5,10)~{\rm TeV}$ (bottom left), $(5,15)~{\rm TeV}$ (bottom right). The blue band denotes the LZ $90\%$ C.L. region for a fully endothermic scenario, while the black curve gives the cross section required to produce one event. This also includes the exothermic contribution that is relevant below $\delta\lesssim100\,\mathrm{keV}$. The red dashed line shows the thermal freeze-out prediction. The colored points reproduce both the observed DM abundance through low-temperature freeze-in and the one-event condition, with the color indicating the corresponding reheating temperature $T_R$.}
    \label{fig:non-leptophobic}
\end{figure}

In the universal coupling scenario, the $Z'$ mass and couplings are constrained by dilepton
resonance searches, $pp\to Z'\to\ell^+\ell^-$, with $\ell=e,\mu$. We
use the latest ATLAS analysis~\cite{ATLAS:2026nqc}, based on
$165~{\rm fb}^{-1}$ of Run~3 data at $\sqrt{s}=13.6~{\rm TeV}$, which
sets $95\%$ CL upper limits on
$\sigma_{\rm fid}(pp\to Z')\,{\rm BR}(Z'\to\ell^+\ell^-)$ for different
values of $\Gamma_{Z'}/m_{Z'}$. We compute the corresponding prediction
with \textsc{MadGraph5\_aMC@NLO}~\cite{Alwall:2014hca} and compare it
with the limit associated with the predicted mediator width. We do not
use the combined Run~2 and Run~3 result, since it is only provided for
specific ATLAS benchmark models.
For all the mediator masses considered, dilepton resonance searches were found to produce no constraints on the cross sections needed to explain the LZ candidate event and reproduce the correct relic abundance.
Let us notice that in the universal coupling model, the dominant decay channel of $\chi_2$ is into neutrinos, which would not induce any signal in the detector.

The numerical results are shown in Fig.~\ref{fig:non-leptophobic}. As in the leptophobic case, varying the
reheating temperature allows the relic-density condition to span a continuous range of
$\sigma_N$. In the universal scenario, we notice that the LZ candidate can be reproduced by the thermal freeze-out for all the benchmark point considered. As in the leptophobic case, lower values of $\delta$,
which would not reproduce the event in the thermal relic scenario, are now
viable with low-reheating freeze-in. 
The viable regions in the reheating temperature are only mildly shifted with
respect to the leptophobic scenario, while a significant shift is seen in $\delta$, as summarized in Table~\ref{tab:LZ_RD_ranges}. Due to the nature of the endothermic scattering there is a kinematic threshold at high $\delta$, limiting the maximum mass splitting reachable. Let us also notice the change on the slope in the one event line is due to non-depletion of the excited state below $\delta\simeq100$ keV. Finally, we note that the LZ 90\% C.L. region (blue band in Fig.~\ref{fig:non-leptophobic}) assumes purely endothermic scattering. We do not include the exothermic contribution to the recast of the C.L. band, since for $\delta \lesssim 100~\mathrm{keV}$ the model would generate additional low-energy recoils, making this region less relevant.

\subsection{Phenomenology of Pseudo-Dirac Dark Matter}

\begin{figure}[h]
\centering
\includegraphics[width=0.98\textwidth]{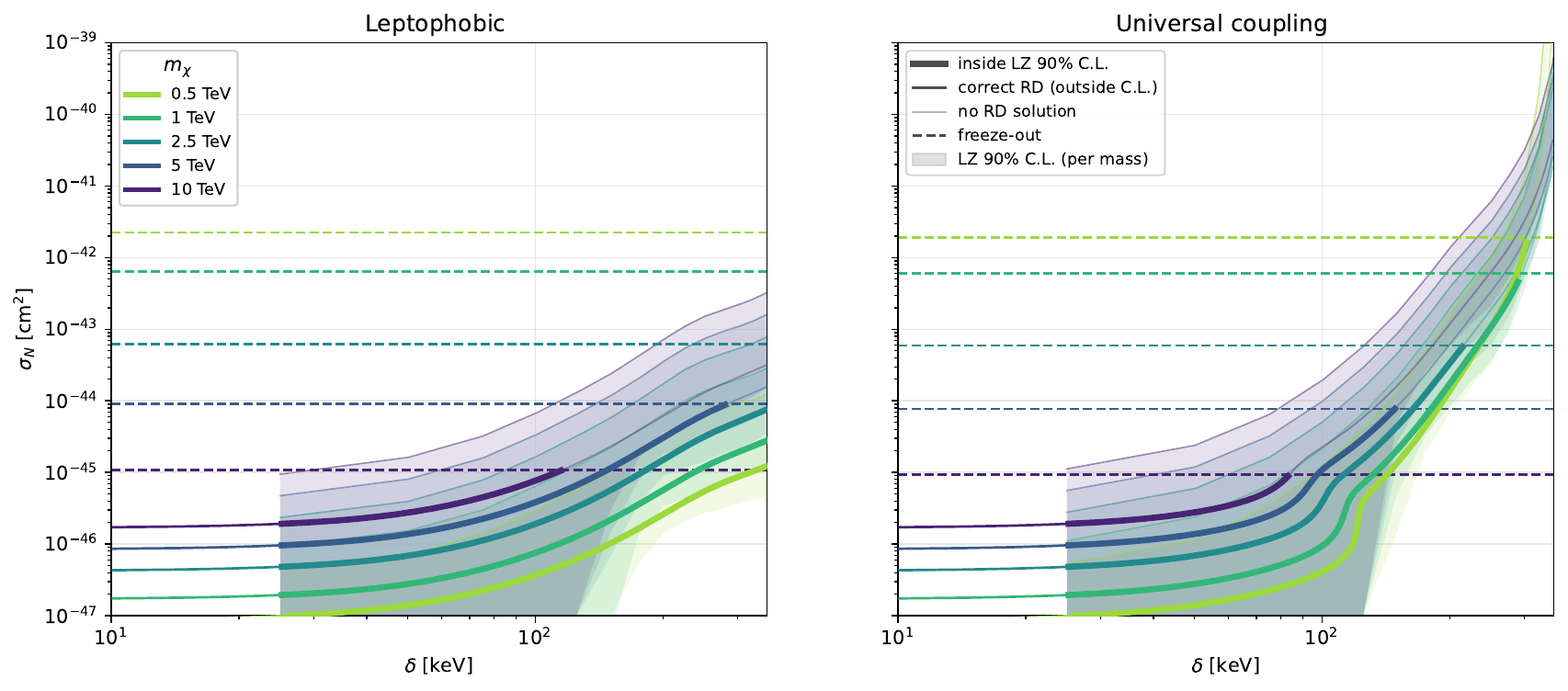}
\caption{One-event lines reproducing the LZ high-energy nuclear-recoil candidate in the $(\delta, \sigma_N)$ plane, for the leptophobic (left) and universal coupling (right) scenarios and several DM masses $m_{\chi_1}$, indicated by the colors. The dashed horizontal lines show the thermal freeze-out prediction, i.e. the thermalization upper bound for the coupling of freeze-in dark matter for each mass. Below this value, the thick lines,  reproduces $\Omega h^2 = 0.12$ in the low-reheating freeze-in  for a suitable $T_R$. The blue band, delimited by the thin black curves, is the LZ 90\% C.L.\ region, for $f = 1/2$ in the leptophobic case~\cite{oliveira2026} and for $\chi_1$ alone in the universal coupling case.}
\label{fig:multimass}
\end{figure}

Fig.~\ref{fig:multimass} shows the relic density curves for $m_{\chi_1} = 0.5$, 1, 2.5, 5 and 10~TeV in the leptophobic (left panel) and universal coupling (right panel) scenario. For all of the benchmark points, part of each curve lies inside the LZ 90\% C.L.\ region. However, the viable window closes as the DM mass increases. 

Two effects drive this behaviour. First, the thermal cross section, $\langle\sigma v\rangle$, is proportional to $m^2_{\chi_1}$. In the contact limit, $\sigma_N$ and $\langle\sigma v\rangle$ depend on the same combination $g_\chi g_q/m_{Z'}^2$, so the relic condition fixes $\sigma_N \propto \mu_N^2 \langle\sigma v\rangle / (N_f\, m_{\chi_1}^2)$, independently of $m_{Z'}$. Since the thermalization condition bounds the freeze-in cross section from above, the upper edge of the splitting moves down, from $\delta \simeq 303$~keV at 0.5~TeV to $\delta \simeq 148$~keV at 5~TeV for the universal case. Second, the DM number density scales as $e^{-m_{\chi_1}/T}$. A heavier candidate therefore needs a larger $\sigma_N$ to produce one event, and the one-event line moves up.

The reheating temperature follows the DM mass. Along the viable segments, $x_R \equiv m_{\chi_1}/T_R$ lies between about $19$ and $27$ for all benchmarks. The interval is narrow because of the exponential factor in Eq.~\eqref{eq:cross_section_final}: a shift of five units in $x_R$ spans four orders of magnitude in $\sigma_N$. At the upper end of $x_R$, $T_R$ approaches the freeze-out temperature and the freeze-in solution merges into the thermal one.

The main difference between the leptophobic and universal coupling scenarios in direct detection arises from the depletion of $\chi_2$. In the leptophobic scenario, the surviving $\chi_2$ population divides the direct-detection phenomenology into the endothermic and exothermic regimes discussed above. The impact on the relic density is comparatively small, the leptonic channels raise the effective final state multiplicity from $N_f = 18$ to $N_f = 22.5$, and the $\sigma_N$ at fixed $T_R$ scale as $1/N_f$, cf.\ Eq.~\eqref{eq:cross_section_final}. The thermalization bound (horizontal dashed lines) is therefore a factor \(1.25\) larger than in the universal-coupling scenario.

The pseudo-Dirac structure also has important consequences for indirect detection. Since the vector interaction is off-diagonal in the Majorana basis, the leading annihilation process is the coannihilation $\chi_1\chi_2\to Z^{\prime *}\to f\bar f$, whereas the tree-level process $\chi_1\chi_1\to f\bar f$ is absent. Consequently, if the excited state $\chi_2$ is efficiently depleted after freeze-out the present-day annihilation rate is suppressed by its residual abundance, allowing the model to evade conventional gamma-ray, cosmic-ray, and CMB constraints. This is indeed the case for the universal coupling scenario, where $\chi_2$ is efficiently depleted by the decay $\chi_2 \to \chi_1 \bar \nu \nu$ in the region of interest. The lifetime written as a function of the nuclear scattering cross section is~\cite{Fitzpatrick:2021cij}
\begin{equation}
    \tau_{\chi_2 \to \chi_1 \nu \bar{\nu}}
\simeq
\frac{540\pi^2 \mu_N^2}{\delta^5 \sigma_N}.
\end{equation}
And we can see from Figure~\ref{fig:non-leptophobic} that we have a relevant residual abundance of $\chi_2$ today for $\delta \lesssim 100$ keV, which is already disfavoured by the nuclear recoil spectra. We will therefore not focus on possible bounds that can arise at low $\delta$. The case of the leptophobic coupling is different because the decay rate $\chi_2 \to \chi_1 +3 \gamma$, is strongly suppressed, with lifetime~\cite{Fitzpatrick:2021cij}
\begin{equation}
    \tau_{\chi_2 \to \chi_1 + 3\gamma}
=
\frac{2.3 \times 10^{15}\, m_e^8 \mu_N^2}
{g_\chi^2 \sigma_N \delta^{13}}
\end{equation}
which does not allow depletion. Notice that such long-lived $\chi_2$ cannot therefore decay inside the detector and introduce spurious signals. Other processes, such as downscattering and self-conversion, can also deplete the excited-state population; however, for the benchmark parameters considered in this work, their rates are too small to lead to efficient depletion \cite{CarrilloGonzalez:2021lxm}, rendering the $\chi_2$ and $\chi_1$ abundance approximately equal from production until the present day. Therefore the coannihilation $\chi_1\chi_2\to Z^{\prime *}\to b\bar b$ could lead to indirect detection signals. However, the upper bound on $\sigma_N$ resides far above the region of interest at values $\mathcal{O}(10^{-41}-10^{-42})$. Nonetheless, the presence of both $\chi_1$ and $\chi_2$ drastically modifies the direct-detection phenomenology, shifting the signal from being dominated by endothermic scattering to being dominated by exothermic scattering, as discussed above. This applies equally to freeze-in and freeze-out production, and therefore alters the viable parameter space compared with analyses in which depletion of the excited state is assumed~\cite{DiMauro:2026ldr}.

In Ref.~\cite{Pospelov:2026ewn} it was shown that solar capture constraints rule out the thermal Higgsino DM model, while the pseudo-Dirac benchmark remains a valid explanation for the LZ candidate~\cite{dimauro2026solarcapturetestsinelastic}. Moreover, it can be shown that the solar capture rate increases with the DM-nucleon cross section and decreases with the DM candidate's mass.  Since the viable cross sections decrease for larger masses, as shown in Fig.~\ref{fig:multimass}, the solar capture rate for any $m_\chi > 1 \,\rm TeV$ is lower than the one obtained for the benchmark in Ref.~\cite{dimauro2026solarcapturetestsinelastic}. Therefore, the pseudo-Dirac model remains a valid explanation for the LZ high-energy nuclear recoil also for low-temperature freeze-in.

Satisfying simultaneously the relic-density condition, the LZ $90\%$ C.L. interval, and the one-event normalization does not by itself constitute a fit to the candidate. These requirements only show that the model can produce an isolated event at the required rate. A viable interpretation should also assign a significant fraction of the predicted signal to recoil energies close to the observed value.

As seen from Table~\ref{tab:LZ_RD_ranges}, only the benchmarks with $m_{\chi_1}=0.5$, $1$, and $2.5~{\rm TeV}$ can reach $E_R^\star\gtrsim200~{\rm keV}$ within the LZ region for the endothermic case. For $m_{\chi_1}=5~{\rm TeV}$, instead, the characteristic recoil energy remains below approximately $150~{\rm keV}$. The signal is therefore expected to be concentrated at lower recoil energies, producing more events below the observed candidate and making this benchmark less suitable for explaining an isolated recoil at $248~{\rm keV}$.

\begin{figure*}[t]
    \centering
    \includegraphics[width=0.78\textwidth]{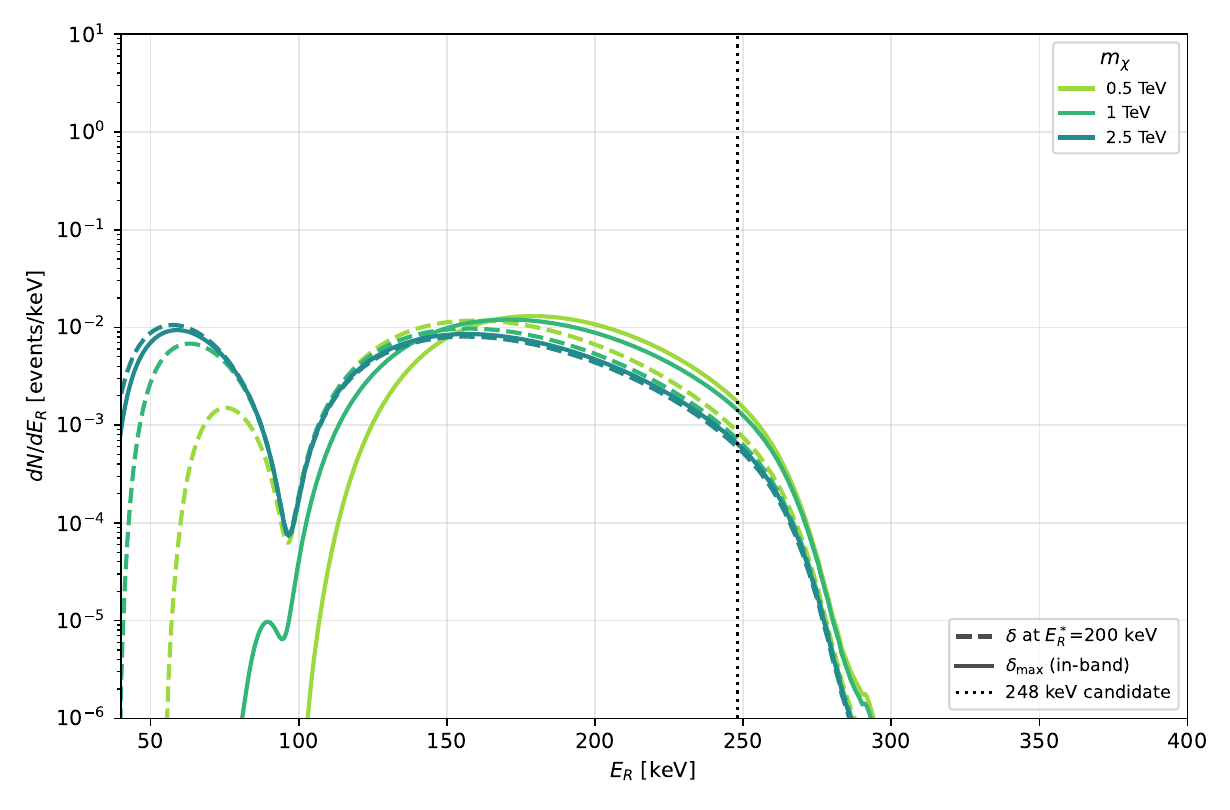}
    \caption{Efficiency-weighted nuclear-recoil spectra for the universal scenario, including the natural xenon isotope abundances and the Helm form factor. The dashed curves correspond to the splitting for which $E_R^\star=200~{\rm keV}$, while the solid curves use the largest splitting lying within the LZ $90\%$ C.L. region. $\delta_{\rm max}$ is the largest value reachable via freeze-in which corresponds approximately to the value of $\delta$ in the freeze-out case. Each point reproduces the observed DM relic abundance through low-temperature freeze-in and satisfies the one-event normalization. The vertical dotted line marks the $248~{\rm keV}$ LZ candidate.}
    \label{fig:recoil_spectra_leptophobic}
\end{figure*}
Since the leptophobic scenario is dominated by exothermic scattering it populates lower recoil energies. Therefore it is less favourable for the interpretation of an isolated event at \(248\) keV. We focus instead on the universal-coupling scenario. Fig.~\ref{fig:recoil_spectra_leptophobic} shows the recoil spectra after including the xenon form factor and the LZ efficiency. The diffraction structure of the Helm form factor separates the spectrum into two recoil regions. For $m_{\chi_1}=2.5~{\rm TeV}$, a sizeable low-energy component remains, while the second maximum lies well below the observed recoil even in the best case scenario $\delta=\delta_{\rm max}$. This benchmark is therefore less favourable for explaining the isolated event. For $m_{\chi_1}=0.5$ and $1~{\rm TeV}$, the larger allowed splittings suppress the low-energy contribution and shift the second recoil region towards the LZ candidate, increasing the relative probability of an event near $248~{\rm keV}$. The candidate nevertheless lies on the falling high-energy side of these spectra, rather than precisely at their maximum.

At the largest allowed splittings, the corresponding minimum velocities are approximately $745$ and $692~{\rm km\,s^{-1}}$ for $m_{\chi_1}=0.5$ and $1~{\rm TeV}$, respectively. Both remain below the mean maximum speed in the detector frame, $v_{\rm esc}+\overline v_E\simeq794~{\rm km\,s^{-1}}$, and are therefore kinematically accessible. The lighter benchmark, however, relies more strongly on the high-velocity tail and is consequently more sensitive to halo uncertainties. We therefore identify the range $m_{\chi_1}\simeq0.5$--$1~{\rm TeV}$ in the universal-coupling scenario as the most favourable among the benchmarks considered. Since larger splittings correspond to smaller reheating temperatures along the simultaneous relic-density and one-event curves, these spectra select the lower part of the $T_R$ intervals reported in Table~\ref{tab:LZ_RD_ranges}.

\section{Conclusions}
\label{sec:concl}

LZ has reported a nuclear-recoil candidate with reconstructed energy \(E_R\simeq248~{\rm keV}\). An elastic spin-independent interpretation is difficult because a signal normalized to this high-energy event would generically predict a much larger number of lower-energy recoils. Endothermic inelastic scattering provides a natural way to suppress this low-energy population and to shift the characteristic recoil energy towards the observed event.

In this work, we have studied this possibility in a pseudo-Dirac fermion model with an off-diagonal vector interaction mediated by a heavy \(Z'\). We considered dark matter production through Boltzmann-suppressed freeze-in at a reheating temperature below the DM mass. In contrast to conventional thermal freeze-out, the relic-density condition is then controlled not only by the interaction strength but also by \(T_R\). For fixed \(m_{\chi_1}\), varying \(T_R\) produces a continuous curve in the \((\sigma_N,\delta)\) plane that reproduces the observed DM abundance. This allows the LZ candidate event to be explained at mass splittings below the corresponding thermal-relic solution.

The cosmological evolution of the heavier pseudo-Dirac state cannot be neglected. \(\chi_1\) and \(\chi_2\) are produced in equal amounts. In the leptophobic scenario, the decay of \(\chi_2\) is highly suppressed and the conversion does not efficiently deplete the excited state. The surviving \(\chi_2\) population gives an exothermic direct-detection component that dominates over the endothermic one and populates lower recoil energies; making an interpretation of an isolated \(248~{\rm keV}\) event less favourable. We notice that, once this effect is included, also the \(m_{\chi_1}\simeq1~{\rm TeV}\) freeze-out interpretation is not viable in the leptophobic realization unless an additional depletion mechanism is present. Low-reheating freeze-in nevertheless is viable albeit disfavoured kinematically.

The universal-coupling scenario has a qualitatively different phenomenology. In the region of interest, \(\chi_2\) can decay through \(\chi_2\to\chi_1\nu\bar{\nu}\), so that the direct-detection signal is predominantly endothermic. The corresponding thermal and low-reheating freeze-in curves can reproduce the LZ candidate for the benchmark masses in the TeV range studied in this work, with the freeze-in extending the viable region towards smaller splittings. 
The recoil spectra provide an additional discriminator beyond the relic-density condition and the one-event normalization. Among the benchmarks considered, the universal-coupling scenario with \(m_{\chi_1}\simeq0.5\)--\(1~{\rm TeV}\) gives the most favourable spectra: the allowed splittings suppress the low-energy contribution and place the high-energy recoil distribution in the spectrum closest to the LZ candidate event. For \(m_{\chi_1}=2.5~{\rm TeV}\), a sizeable low-energy component remains, while for heavier masses the largest accessible characteristic recoil energies fall below the observed event. For the viable scenario of $m_{\chi_1}=1~{\rm TeV}$, the reheating temperature is predicted to be in the range $T_R\simeq39-47~{\rm GeV}$. 

These results show that the interpretation of the LZ high-energy candidate depends jointly on the particle-physics realization and on the cosmological history. Moreover, in low-reheating freeze-in scenarios, direct detection can become a probe of the pre-BBN thermal history, with the measured scattering rate constraining the reheating temperature to a narrow range. Since our analysis uses a one-event normalization rather than a full event-level likelihood, a dedicated statistical analysis of the recoil spectrum, including the two-state halo where relevant, will be required to establish whether this scenario provides a decisive explanation of the LZ candidate.

\section*{Acknowledgement}
The authors thank Oleg Lebedev for the fruitful and enlightening discussions and Katelin Schutz for drawing our attention to the possible survival of the excited state. FC acknowledge partial financial support from the Research Ireland Awards Grant 21/PATH-S/9475 (MOREHIGGS) under the SFI-IRC Pathway Program. DCA acknowledge funding from the European Union’s Horizon Europe research and innovation programme under the Marie Skłodowska-Curie Staff Exchange grant agreement No 101086085 - ASYMMETRY, and thanks Lawrence Berkeley National Laboratory for its hospitality during the development of this work. VO acknowledges the financial support from FCT through the doctoral program grant with the reference PRT/BD/154629/2022 (\url{https://doi.org/10.54499/PRT/BD/154629/2022}).\\

\clearpage
\clearpage
\bibliographystyle{JHEP}
\bibliography{biblio.bib}

\end{document}